\documentclass[sigconf]{acmart}
\usepackage{array}
\usepackage{multirow}
\usepackage{enumitem}
\usepackage{dblfloatfix}
\usepackage{svg}
\usepackage{xcolor}
\usepackage{soul}

\AtBeginDocument{%
  \providecommand\BibTeX{{%
    \normalfont B\kern-0.5em{\scshape i\kern-0.25em b}\kern-0.8em\TeX}}}

\copyrightyear{2023} 
\acmYear{2023} 
\setcopyright{rightsretained} 
\acmConference[CHI '23]{Proceedings of the 2023 CHI Conference on Human Factors in Computing Systems}{April 23--28, 2023}{Hamburg, Germany}
\acmBooktitle{Proceedings of the 2023 CHI Conference on Human Factors in Computing Systems (CHI '23), April 23--28, 2023, Hamburg, Germany}\acmDOI{10.1145/3544548.3581308}
\acmISBN{978-1-4503-9421-5/23/04}

\definecolor{mygray}{gray}{0.25}
\definecolor{mygray2}{gray}{0.8}
\definecolor{mygray3}{gray}{0.9}
\definecolor{amber}{rgb}{1.0, 0.75, 0.0}
\definecolor{mygreen}{rgb}{0.0, 0.5, 0.0}

\usepackage{acmart-taps}

\aptLtoXcmd{}{\newenvironment{myquote}%
  {\list{}{\leftmargin=0.1in\rightmargin=0.1in}\item[]}%
  {\endlist}}

\begin{document}

\title[Understanding Street-Level Complexities within the CWS Decision-Making Ecosystem]{Rethinking "Risk" in Algorithmic Systems Through A Computational Narrative Analysis of Casenotes in Child-Welfare}

\author{Devansh Saxena}
\affiliation{%
  \institution{Marquette University}
  \streetaddress{Cudahy Hall, 1313 W Wisconsin Avenue}
  \city{Milwaukee}
  \state{WI}
  \postcode{53233}
  \country{USA}}
\email{devansh.saxena@marquette.edu}

\author{Erina Seh-Young Moon}
\affiliation{%
  \institution{University of Toronto}
  \streetaddress{140 St. George Street}
  \city{Toronto}
  \state{Ontario}
  \country{Canada}}
 \email{erina.moon@mail.utoronto.ca}

\author{Aryan Chaurasia}
\affiliation{%
  \institution{University of Toronto}
  \streetaddress{140 St. George Street}
  \city{Toronto}
  \state{Ontario}
  \country{Canada}}
  \email{aryan.chaurasia@mail.utoronto.ca}

\author{Yixin Guan}
\affiliation{%
  \institution{University of Toronto}
  \streetaddress{140 St. George Street}
  \city{Toronto}
  \state{Ontario}
  \country{Canada}}
  \email{yixin.guan@mail.utoronto.ca}  

\author{Shion Guha}
\affiliation{%
  \institution{University of Toronto}
  \streetaddress{140 St. George Street}
  \city{Toronto}
  \state{Ontario}
  \country{Canada}}
 \email{shion.guha@utoronto.ca}

\renewcommand{\shortauthors}{Devansh Saxena et al.}

\begin{abstract}
Risk assessment algorithms are being adopted by public sector agencies to make high-stakes decisions about human lives. Algorithms model “risk” based on individual client characteristics to identify clients most in need. However, this understanding of risk is primarily based on easily quantifiable risk factors that present an incomplete and biased perspective of clients. We conducted a computational narrative analysis of child-welfare casenotes and draw attention to deeper systemic risk factors that are hard to quantify but directly impact families and street-level decision-making. We found that beyond individual risk factors, the system itself poses a significant amount of risk where parents are over-surveilled by caseworkers and lack agency in decision-making processes. We also problematize the notion of risk as a static construct by highlighting the temporality and mediating effects of different risk, protective, systemic, and procedural factors. Finally, we draw caution against using casenotes in NLP-based systems by unpacking their limitations and biases embedded within them.
\end{abstract}



\maketitle

\section{Introduction}
Public sector agencies such as the child-welfare, criminal justice, unemployment services, and public education have experienced a fundamental economic shift over the last two decades in regard to how governance practices are carried out and how clients are “assisted” by street-level civil servants. Economic principles centered in cost reduction, efficiency, and productivity are now being applied to public services where several sectors have experienced privatization with a core focus on optimization and austerity \cite{eubanks2018automating, redden2020datafied}. “Risk” has been one of the core organizing principles in this economic shift in governance where administrative data accumulated by government agencies about citizens purportedly allows them to preemptively recognize clients in the riskiest circumstances \cite{callahan2018paradox, andrejevic2019automating}, i.e. – clients most in need of public assistance, clients most likely to harm others or engage in unlawful behavior, and clients who pose the most risk to governmental apparatus in terms of resources used. Government agencies are employing algorithmic systems to predict outcomes such as the risk of recidivism \cite{greene2020hidden, dressel2018accuracy}, risk of child maltreatment \cite{saxena2020human, cheng2022child}, risk of long-term unemployment \cite{holten2020shifting, ammitzboll2021street}, risk of extended homelessness \cite{eubanks2018automating}, among others. This preemptive recognition and mitigation of "risk" through predictive models is a defining characteristic of what scholars have called \textit{digital era governance} \cite{busch2018digital} or \textit{digital welfare states} \cite{alston2019report}. However, there is a mismatch between how risk is quantified \textit{empirically} \cite{saxena2020human} based on administrative data versus how it is understood \textit{theoretically} \cite{austin2020risk} in these public sector domains. 

Empirical risk predictions hold the promise of providing consistent, cost-effective, and objective decisions, and bringing a new data-driven perspective to government agencies where data would bear the promise of future bureaucratic efficiencies \cite{holten2020shifting}; however, audits of these systems have revealed that they instead achieve worse outcomes \cite{saxena2021framework2}, embed human biases present in administrative data \cite{dressel2018accuracy, tan2018investigating}, appear nonsensical to workers \cite{kawakami2022improving, saxena2021framework2}, and exacerbate existing racial biases \cite{cheng2022child}. Consequently, researchers studying fairness, accountability, and transparency in algorithmic systems have developed technical definitions of "fairness" and "bias" and formulated them into systems design to achieve equitable outcomes. These approaches may lead to the development of systems that are mathematically fair, however, they still continue to focus on a narrow understanding of "risk" as derived from empirical administrative data while drawing attention (and resources) away from the complexities in the decision-making ecosystem and ecological nature of risk that families and caseworkers experience on the street-level \cite{saxena2020human}. To address these gaps in computational research, SIGCHI scholars have begun to examine how quantitative methods can be used to uncover complexities and latent patterns within sociotechnical systems \cite{cambo2022model, gordon2022jury, saxena2022unpacking}. In this study, we accept this call, and instead of quantifying risks using administrative data, we focus on uncovering human interactions between caseworkers, families, and other child-welfare stakeholders to understand the multiplicity and temporality of risk factors that arise in child-welfare cases through the lens of computational narrative analysis of casenotes, i.e. - rethinking "risks" as they occur on the street-level and recorded in caseworkers' narratives as opposed to what is quantified in the administrative data. We used the socioecological model of health and development that has been recently used to study risk, protective, systemic, and procedural factors associated with child maltreatment \cite{austin2020risk} as the theoretical lens for grounding our quantitative analysis. In this study, we pose the following overarching research questions: 

\begin {itemize}\leftskip-16pt
  \item \textcolor{mygray}{\textbf{RQ1:} Which factors in the child-welfare ecosystem directly impact street-level decision-making and family well-being?}
  \item \textcolor{mygray}{\textbf{RQ2:} How do these critical factors interact and what is the impact of this interplay on decision-making?}
  \item \textcolor{mygray}{\textbf{RQ3:} How do these critical factors fluctuate and mediate each other throughout the life of child-welfare cases?}
\end{itemize}

Abebe et al. \cite{abebe2020roles} argue that computational research has meaningful roles to play in addressing social problems by highlighting deeper patterns of injustice and inequality. In this regard, they formulate roles that computing can play and define \textbf{\textit{computing as rebuttal}} when it illuminates the boundaries of what is technically feasible and define \textbf{\textit{computing as synecdoche}} when it makes long-standing social problems newly salient in the public eye. In this study, we assume these roles and make the following contributions:

\begin {itemize}\leftskip-16pt
  \item We use computational narrative analysis \cite{saxena2022unpacking, antoniak2019narrative} to uncover the different risk, protective, systemic, and procedural factors that impact street-level decision-making and draw these connections to the theoretical understanding of risk in sociology and child-welfare literature \cite{austin2020risk}.
  \item We showcase how these different factors interact with each other where systemic and procedural factors can amplify the risks that families experience in the child-welfare system. 
  \item We highlight how these factors change over time and can compound uncertainty in decision-making due to a lack of clarity about the trajectory of cases. We further complicate the use of predictive risk models (PRMs) because no temporal point estimate of risk offers a complete picture of family well-being.
  \item We surface the limitations and biases embedded within child-welfare casenotes and draw caution against using these narratives for downstream tasks (e.g., predicting the risk of child maltreatment) in NLP-based systems. Alternately, an upstream approach, as adopted by this study can help uncover dynamic and transitory signals embedded within the sociotechnical practices of decision-making.
\end{itemize}

This study responds to calls within SIGCHI research to investigate complex sociotechnical systems from both a qualitative and quantitative lens to understand the opportunities and limitations of computational research towards highlighting social problems and addressing injustices \cite{aragon2022human, mlforsociology, abebe2020roles}.


\section{Related Work}
In this section, we first discuss recent research within SIGCHI conducted in the public sector followed by research conducted on computational text analysis of sociotechnical systems.

\subsection{Public Sector Research within SIGCHI}
The SIGCHI community has a long-standing history of conducting research in the public sector and designing sociotechnical systems that empower public sector workers \cite{holten2020shifting, ammitzboll2021street, saxena2020conducting, kawakami2022care} and affected communities \cite{stapleton2022has, brown2019toward, stapleton2022imagining, badillo2018chibest}. Most relevant to this study, SIGCHI research spans across digital civics \cite{dow2018between, lodato2018institutional}, digital governance \cite{busch2018digital, light2019breakdown}, and algorithmic governance \cite{kawakami2022improving, robertson2020if} where HCI scholars have studied issues of citizen engagement in the public space \cite{koeman2015everyone, dow2018between}, citizen activism \cite{massung2013using}, empowerment of affected communities \cite{brown2019toward, pierre2021getting}, centering worker well-being in gig work \cite{zhang2022algorithmic, toxtli2021quantifying, toxtli2020reputation, savage2021research}, and engaging in action research with community partners \cite{cooper2022systematic}. As government agencies experience renewed neoliberal market forces centered in austerity and privatization \cite{legreid2017transcending, young2019artificial}, digital governance platforms are being developed through public-private partnerships \cite{county2017developing} or by contracting tech startups \cite{sas, pap}. Here, HCI researchers have also questioned the forms and limitations of participatory design in the public sector that is increasingly experiencing the deployment of technologies developed through public-private partnerships for the administration of smart cities \cite{lodato2018institutional, dow2018between}. In addition, HCI scholars have also brought into question the core function of government services that were designed to act as "caring platforms" by serving the public good but are now being operated based on business models of private corporations \cite{light2019breakdown}. That is, public services designed to “serve” the people should not be optimized or reduced to the performance metrics of the business world. To oppose this, HCI scholars have also advocated for adopting “care” as a design framework for developing systems that upload values of a caring democracy \cite{meng2019collaborative, toombs2018designing, golsteijn2016sens}. Here, a critical aspect of civil servants’ labor involves conducting care work in the context of risk. Gale et al. \cite{gale2016towards} describe this as “risk work” where civil servants are tasked with assessing and managing risks, minimizing risks in practice, and translating risk in different contexts. However, risk work (i.e., human discretionary work) in the public sector such as assessing the risk of child maltreatment {\cite{saxena2020human}}, risk of recidivism {\cite{greene2020hidden}}, risk of long-term unemployment {\cite{ammitzboll2021street}}, risk of long-term homelessness {\cite{Kithulgoda2022}}, etc. is increasingly mediated through algorithmic systems.

Consequently, HCI researchers have also started studying how human discretionary work is changing in the public sector and adopted Lipsky's theory of street-level bureaucracy {\cite{lipsky2010street}} to understand how street-level bureaucrats or civil servants (e.g., caseworkers, police officers, judges, educators) reflexively balance the needs of citizens against the demands of policymakers. With the adoption of digital technologies and several decisions about citizens being made from `behind a screen', Bovens and Zouridis adopted Lipsky's theory to highlight how public services were transforming into screen-level bureaucracies {\cite{bovens2002street}}. Most recently, Alkhatib and Bernstein adapted Lipsky's theory into \textit{street-level algorithms} {\cite{alkhatib2019street}} to further highlight the shift in governance as a result of algorithmic decision-making. This has further allowed researchers to investigate the intersection of human discretionary work conducted at the street-level and algorithmic decision-making \cite{paakkonen2020bureaucracy, saxena2021framework2, veale2018fairness, holten2020shifting, ammitzboll2021street, robertson2021modeling, saxena2020child}. Much of this scholarly work conducted in these domains has found that there is a mismatch between how risk is empirically quantified and predicted by algorithms versus how risk is theoretically understood, informs street-level practices, and impacts families in need of public services. This mismatch also leads to unreliable decision-making and frustrations on part of civil servants who are mandated to use algorithmic systems \cite{kawakami2022improving, saxena2021framework2}. Specific to the child-welfare system, algorithmic governance systems in the form of predictive risk models (PRMs) are being adopted as a means to proactively recognize cases where children are at high risk for maltreatment and offer targeted services to these families. However, recent studies have found that such systems exacerbate racial discrimination and inequalities and further undermine the rights of low-income communities \cite{cheng2022child, parker2022examining}. 

A nationwide survey on predictive analytics in child-welfare conducted by the American Civil Liberties Union (ACLU) in 2021 revealed that 26 states have considered employing predictive analytics in child-welfare \cite{samant2021family}. Of these 26 states, 11 are currently using them \cite{samant2021family}; however, audits of these systems reveal that they are achieving worse outcomes for families and exacerbating racial biases \cite{lacounty2017, chicago2017, oregon2022, allegheny2022}. Due to these concerns, Los Angeles County and Illinois have shut down their predictive analytics programs in the past \cite{lacounty2017, chicago2017} with Oregon recently joining their ranks in June 2022 \cite{oregon2022}. A recent study conducted by Cheng and Stapleton et al. \cite{cheng2022child} on the Allegheny Family Screening Tool (AFST) found that AFST-predicted decisions were racially biased, and workers reduced these biases by overriding erroneous decisions. AFST was designed to mitigate call screeners' biases and subjective decisions and augment decision-making by making it more objective through data. Ironically, AFST has introduced more complexities in decision-making and the call screeners are the ones mitigating algorithmic biases. Another recent study conducted on Eckerd Rapid Safety Feedback showed that the algorithm did not reduce incidences of subsequent child maltreatment \cite{parker2022examining}. A literature review of child-welfare algorithms in the United States also revealed other sources of biases embedded in the predictors, outcomes, and computational methods being used to develop these systems \cite{saxena2020human}. This study also highlights that the majority of algorithms used in CWS are predictive risk models. Finally, a study conducted by Kawakami et al. \cite{kawakami2022improving} on AFST showed that there were misalignments between AFST’s predictive target and call screeners' decision-making objective where call screeners relied more on their contextual understanding of the family and risk factors to make decisions rather than empirical risk as predicted by AFST. That is, call screeners, focused more on contextual risk factors that families experienced on the street-level as opposed to risk quantified using administrative data.

Federal initiatives such as improved data infrastructures for CWS \cite{harrison2018tale} have paved the way for tech startups to develop and pitch algorithmic systems to human services agencies across different states \cite{eckerd, pap, mind-share, sas}. However, there is a need to critically examine the current points of failure in the design of predictive risk models (PRMs). Critical to the conversation about PRMs is also the underlying principle of "risk" and how its understanding has shifted in response to the restructuring of public services to be economically efficient and productive \cite{callahan2018paradox, redden2020datafied, andrejevic2019automating}. Traditionally, child-welfare services have focused on risks and protective factors within families to be able to provide them with individualized care. However, with a shift towards an empirical understanding of risk and the introduction of PRMs, risk has now become a function of client characteristics as existing in prior cases and their impact on a predictive outcome (i.e., risk of maltreatment). That is, risk is estimated based on historical administrative data and is being used to identify the "deserving poor" who pose the most risk to governmental apparatus \cite{eubanks2018automating}. Redden et al. \cite{redden2020datafied} refer to this as the embedded logic of actuarialism that also obfuscates and drives attention away from social and structural issues that bring poor and vulnerable communities under the attention of public services such as the child-welfare system \cite{keddell2015ethics}.

\subsection{Computational Text Analysis Research within SIGCHI}
Recent works studying sociotechnical systems have employed computational text analysis techniques such as topic modeling, sentiment analysis, and part-of-speech tagging to understand sociotechnical systems \cite{saxena2022unpacking, birthstories2019, thighgap2016}. Nguyen et al. \cite{howwedo2020} argue computational text analysis on texts involves unpacking textual information that is inherently socially and culturally situated where there exists no absolute ground truth. While this poses challenges, this method also offers opportunities to uncover dynamic and transitory phenomena present in sociotechnical systems \cite{ackerman2000}. Prior research has shown that computational text analysis can aid traditional qualitative methods by quickly scaling to large text corpora, aggregating text for analysis, and reducing directionality biases or qualitative oversimplifications \cite{jockers13, realityandml2020, howwedo2020}. Furthermore, recent research has found that machine learning techniques such as topic modeling carry similarities with qualitative methods such as grounded theory, and offer supporting and complementary insights into text \cite{muller2016machine}. For example, Baumer et al. \cite{baumer2017comparing} employed topic modeling and grounded theory on survey responses and found that the two methods yielded similar results, although the former uncovered patterns at lower abstraction levels. 

Various domains, including public policy, child welfare, health, and communication, have applied computational text analysis on varying lengths of texts to investigate issues relevant to the public sector \cite{whyileft2020, tmforqualpolicy, saxena2022unpacking, realityandml2020, birthstories2019}. Notably, Saxena et al. \cite{saxena2022unpacking}, and Antoniak et al. \cite{birthstories2019} found that topic modeling and sentiment analysis on dense and unstructured narrative texts can provide insights not necessarily revealed via manual qualitative methods. Through topic modeling, they showed the technique could uncover invisible patterns of human activity, constraints that affect human decision-making within the domain, and latent day-to-day power dynamics between agents. In a similar vein of work, Abebe et al. \cite{maternalmortality2020} found that computational text analysis of texts can uncover holistic and contextualized details in pregnancy-related tweets and could predict maternal mortality rates at a higher accuracy rate than using socioeconomic and risk variables. Prior applications of topic modeling have also found evidence showing how the technique can support manual analyses of text. For example, Rodriguez and Storer \cite{whyileft2020} showed that by plotting a topic model correlation network for tweets related to domestic violence, topic modeling can provide a descriptive analysis of texts, which is comparable to first-round qualitative analysis. Isoaho et al. \cite{tmforqualpolicy} also noted policy analysis journals extensively use topic modeling as a computational text analysis technique because it can aid manual analyses of texts. Most recently, Showkat and Baumer {\cite{showkat2022s}} engaged in speculative design workshops with journalists and legal experts and examined these domain experts' value expectations regarding automated NLP systems. Their study uncovered tensions around the technical implementation of such systems and implications for when 'not-to-design' them.

Public administrative work often involves collaborative decision-making where civil servants continuously negotiate with multiple stakeholders involved in cases and document case-related information drawn from multiple sources (e.g., meeting notes with other caseworkers, observations, and email or phone exchanges) \cite{ammitzboll2021street, saxena2022unpacking}. While this sector frequently uses predictive algorithms, civil servants have expressed doubt on the utility of such technologies \cite{ammitzboll2021street, feng22}. Instead, civil servants have expressed a desire for technology to support work processes and case management rather than profiling individuals \cite{holten2020shifting}. Responding to these stakeholder needs; we applied computational narrative analysis to uncover critical aspects of child-welfare to better understand the domain. For this study, we applied CorEx \cite{corex}, a semi-supervised topic model which can be used to uncover topics that are specifically associated with the above factors. Unlike unsupervised LDA topic models, which are prone to highlighting dominant themes in texts, CorEx incorporates user-provided domain knowledge in the form of anchor words that allow the topic model to uncover specific topics of interest associated with these anchor words \cite{arefieva21, rizvi19, nizzoli20}. 

\vspace{-0.2cm}
\section{Research Context}
We partnered with a child-welfare agency in a metropolitan area in a Midwestern U.S. state that is part of the broader child-welfare system that was recently investigated by Saxena et al. (2021) \cite{saxena2021framework2}. This agency is contracted by the state’s Department of Children and Families (DCF) and provides child-welfare services to families that are currently under investigation by DCF. Allegations of child maltreatment are investigated by DCF’s Initial Assessment (IA) caseworkers and if maltreatment is substantiated and the case is opened for a CPS investigation, the family is then referred to this non-profit agency to provide child-welfare services. The agency must comply with all DCF standards and policies and meet its accountability requirements. During the initial court hearing, mandatory services and supervised visitation requirements are negotiated between each parent’s attorney, the district attorney’s office, and the judge. The agency provides case management services, parenting classes, permanency consultations, services to foster youth transitioning into adulthood, in-home services when children are not removed from the care of parents, foster care and adoption services, and family preservation services. It is important to note here that critical decision-making power in regard to reunification, termination of parental rights, transfer of guardianship, and adoption sits with the court system and caseworkers can only make recommendations to the district attorney’s office. We obtained Institutional Review Board (IRB) approval from our mid-sized private research university to use casenotes for this research.


Critical to the understanding of child-welfare is also the \textbf{Adoption and Safe Families Act (1997).} This legislation introduced some of the most sweeping changes to the child-welfare system and shifted the focus primarily toward child safety concerns and away from the policy of reuniting children with parents regardless of prior neglect/abuse. It introduced federal funding to assist states with foster care, adoption, and guardianship assistance and expanded family preservation services. In addition, it also introduced a 15-month timeline where states must proceed with the termination of parental rights if the child has been in foster care for 15 out of the last 22 months \cite{gossett2017client}. This speedy termination of parental rights has received widespread criticism but still establishes the restrictive legislative framework within which caseworkers must conduct their work \cite{guggenheim2021racial}. To ensure expedited permanency\footnote{Permanency is defined as reunification with birth parents, adoption or legal guardianship.} for foster children, the agency employs concurrent planning such that two simultaneous plans begin when a child enters foster care -- a plan for reunification with birth parents and a plan for adoption or transfer of guardianship if reunification is not possible (henceforth, \textit{permanency plan}). The goal here is to ensure that children do not incessantly stay in foster care if reunification fails because extended stay and interaction with CWS lead to poor long-term outcomes for foster children where they are unable to form lasting relationships.

Saxena et al. (2022) {\cite{saxena2022unpacking}} used unsupervised LDA topic models to study casenotes and uncovered patterns of invisible labor undertaken by caseworkers as well as showed how different systemic constraints impacted different families based on case complexity and their level of need. Their study conducted the first computational inspection of child-welfare casenotes and provided the computational basis for conducting similar studies in the public sector that seek to uncover latent contextual signals embedded in these sociotechnical systems. In this study, we go a step further and focus on the dynamic and transitory factors that impact caseworkers' decision-making and family well-being. We use semi-supervised topic models {\cite{gallagher2017anchored}} that embed domain knowledge in the form of anchor words to specifically uncover different risk, protective, systemic, and procedural factors that impact street-level decision-making. By embedding domain knowledge based on Austin et al.'s framework {\cite{austin2020risk}} into the CorEx topic model, we are able to guide the model towards specific topics of interest and uncover the multiplicity and temporality of risk factors that are experienced by families and impact caseworkers' decision-making. We further problematize the notion of empirical risk by highlighting the various systemic and procedural factors that augment risks posed to families but can not be quantified.


\section{Methods}
In this section, we first introduce the casenotes dataset and the data cleaning process. Next, we discuss the data analysis and interpretation process. For this study, we adopted Correlation Explanation (CoRex), a semi-supervised topic modeling method developed by Gallagher et al. \cite{gallagher2017anchored} that allows us to incorporate existing domain knowledge into the topic generation process via the use of anchor words. Unlike the generative topic modeling approach (i.e., LDA topic models) employed by Saxena et al. (2022) {\cite{saxena2022unpacking}} which requires specifications for hyperparameters and detailed assumptions, this study uses semi-supervised CoRex topic models that do not assume an underlying generative model. CoRex allows us to embed domain knowledge through anchor words which further promote topic separability and representation. In addition, generative topic models may only portray dominant themes (or topics) in a corpus, however, CoRex, through the incorporation of meaningful domain words, allows us to surface topics that may otherwise be underrepresented in the corpus.
 
\subsection{Dataset}
We obtained casenotes written by the Family Preservation Services (FPS) team. FPS focuses on assisting birth parents achieve reunification with their children by providing crisis support, parenting classes, and helping improve family functioning \cite{fps_online}. This team works closely with families throughout the child-welfare process and interacts with them in-person on a regular basis. The success and effectiveness of FPS is assessed in terms of how many families are successfully reunified. Here, casenotes serve multiple purposes – 1) they provide a roadmap of all interactions and decisions made and are submitted to the DA’s office if/when FPS recommends reunification for a family, 2) they highlight birth parent's progress in their efforts to achieve reunification, 3) they ensure accountability among all caseworkers (i.e., family preservation team and case management team) and consistent recording of interactions \cite{cpsmanual, sidell15}. Writing detailed casenotes is a central component of FPS caseworker duties who are mandated to follow documentation standards established at the agency \cite{cpsmanual}. We manually analyzed several sources of text data such as family assessments, safety plans, and discharge summaries, and settled on FPS casenotes for this study since they carried the most detailed and contextual information from ongoing face-to-face interactions with families as compared to the casenotes of investigative or initial assessment (IA) caseworkers that contained 'perceived' risks from initial interactions. That is, we conducted a significant amount of manual exploration to assess which data sources were useful and appropriate for analysis. We obtained records of 12,391 casenote entries (the ‘dataset’) for 462 families referred to the agency around May 1, 2019, and worked with Family Preservation until December 31, 2021, or were discharged sooner.

\subsection{Data Cleaning, Preparation, and Anonymization}
The dataset contains casenote entries for families identifiable by their unique family identification numbers. To understand the dynamic relationship between CW staff and family members, we compiled the narrative casenotes for each of the 462 families by their family identifier (i.e., the `family ID') in chronological order. We then cleaned and anonymized the casenotes by removing punctuation and names if they appeared in the 2010 U.S. Census ad Social Security names dataset \cite{census2010surname, ssbabynamesWI}. Numbers in the texts were also replaced with the label \textit{NUM} to prevent numbers from raising confounding signals in our analysis. Lastly, consistent with Schofield et al. \cite{schofield2017pulling} who found removing stopwords led to superficial improvements in topic model solutions, we kept all short words and stop words in the texts as we found regardless of whether we removed these words, the topic models yielded no significant variations. Table \ref{tab:statistics} depicts the summary corpus statistics after we followed the above cleaning and preprocessing steps. 

\begin{table} 
\small
\centering      
\begin{tabular}{c | c}  
\hline         
\textbf{Metric} & \textbf{Value} \\ [0.5ex] 
\hline
Number of casenotes with more than 1500 words & 134  \\   
Average number of words per casenote & 1,461  \\ 
Number of words in longest casenote & 16,601  \\ 
Number of unique words & 20,751 \\ [1ex]
\hline      
\end{tabular}
\caption{Corpus Statistics} 
\label{tab:statistics}  
\vspace{-0.7cm}
\end{table}

\subsection{Data Analysis Approach}
In this section, we discuss our data analysis approach for our three research questions. Saxena et al. (2022) \cite{saxena2022unpacking} showed that Latent Dirichlet Allocation (LDA) can be an effective method to computationally study casenotes in the child-welfare system. However, LDA is an unsupervised generative probabilistic method \cite{jelodar2019latent}, which does not have the option to incorporate domain knowledge in the modeling and topic generation process. As such, we build upon this prior study by adopting a semi-supervised Correlation Explanation (CorEx) topic modeling approach. This approach uses word-level domain knowledge by embedding anchor words. According to Gallagher et al. \cite{gallagher2017anchored}, anchored CorEx can offer the following advantages compared to LDA methods: 1) anchoring words allows for topic separability. The topic clusters generated by anchored CorEx have been found to be more homogeneous and contain adjusted mutual information, 2) Anchored CorEx can represent topics better. Anchoring domain knowledge to a single topic can help uncover representative topics, and 3) anchored CorEx allows the user to explore complex issues within a document by finding interesting and non-trivial aspects within the texts. 

\subsubsection{CorEx Topic Modeling (RQ1)}
\label{section:CorEx}
To answer RQ1 and inform the selection of anchoring words for analysis, we picked 10 families that had 10-15 interactions with the agency, another 10 families that had 30-35 interactions, and finally, 10 families with the most interactions and manually inspected their casenotes to understand the factors that impacted critical decisions and family well-being. A word map was made to facilitate our examination. Next, we used the socioecological model of health and development that has been recently used to study risk, protective, systemic, and procedural factors associated with child maltreatment \cite{austin2020risk} as our theoretical lens to be able to incorporate domain knowledge into our quantitative analysis (in the form of anchor words). Risk factors refer to parental experiences, behaviors, and characteristics that increase the likelihood of maltreatment (e.g., mental health, drug use, domestic violence) \cite{austin2020risk}. Protective factors are characteristics that mediate risk factors and reduce the likelihood of maltreatment (e.g., social support system, self-regulation, social skills). Systemic factors (or environmental/community factors) refer to socioeconomic factors such as employment, housing, health insurance, and transportation that impact low-income families. Finally, procedural factors (or societal factors) refer to the policies, protocols, and street-level regulations that underscore the entire child-welfare process and must be followed by families, caseworkers, and all other involved parties, i.e. - procedural factors establish the legislative framework (or the constraints) within which all the decisions must be made.

Therefore, we specified anchoring words for four topics based on these critical factors from the socioecological model \cite{austin2020risk} and also our manual inspection of casenotes. The anchoring words for the four topics are shown in table \ref{tab:anchorwords}. To select the optimal model for our data, we tried to maximize the Total Correlation (TC) value of the models. Total correlation measures the total dependence of topics on the document. The higher the TC value, the more effective the model is in describing the document. We also considered two other aspects, the number of topics and the anchor strength. The anchor strength controls how much weight CorEx puts toward maximizing the mutual information between the anchor words and their respective topics. Anchoring strength is positively correlated with TC. Gallagher et al. \cite{gallagher2017anchored} suggest that setting anchor strength from 1.5-3 can nudge the topic model towards the anchor words and setting it to a value greater than 5 can strongly enforce the CorEx topic model to find topics associated with the anchor words. In our analysis, we found that TC in the CorEx model tends to increase as the number of topics increases. For interpretability, we limited the number of topics to under 20. In the model selection process, we ran all combinations with the topic number from 4 to 20 and the anchor strength from 1 to 6. The model with an anchor strength of 6 and topic number of 19 showed maximum TC. We, therefore, decided on these parameters for the final model for our analysis. Next, four co-authors of this paper individually interpreted and labeled topics based on top keywords and exemplar casenotes. Then, the authors discussed the interpretations and refined topic labels until all authors reached a consensus.

\subsubsection{Qualitative Axial Coding (RQ2)} 
\begin{figure*}
  \includegraphics[scale=0.40]{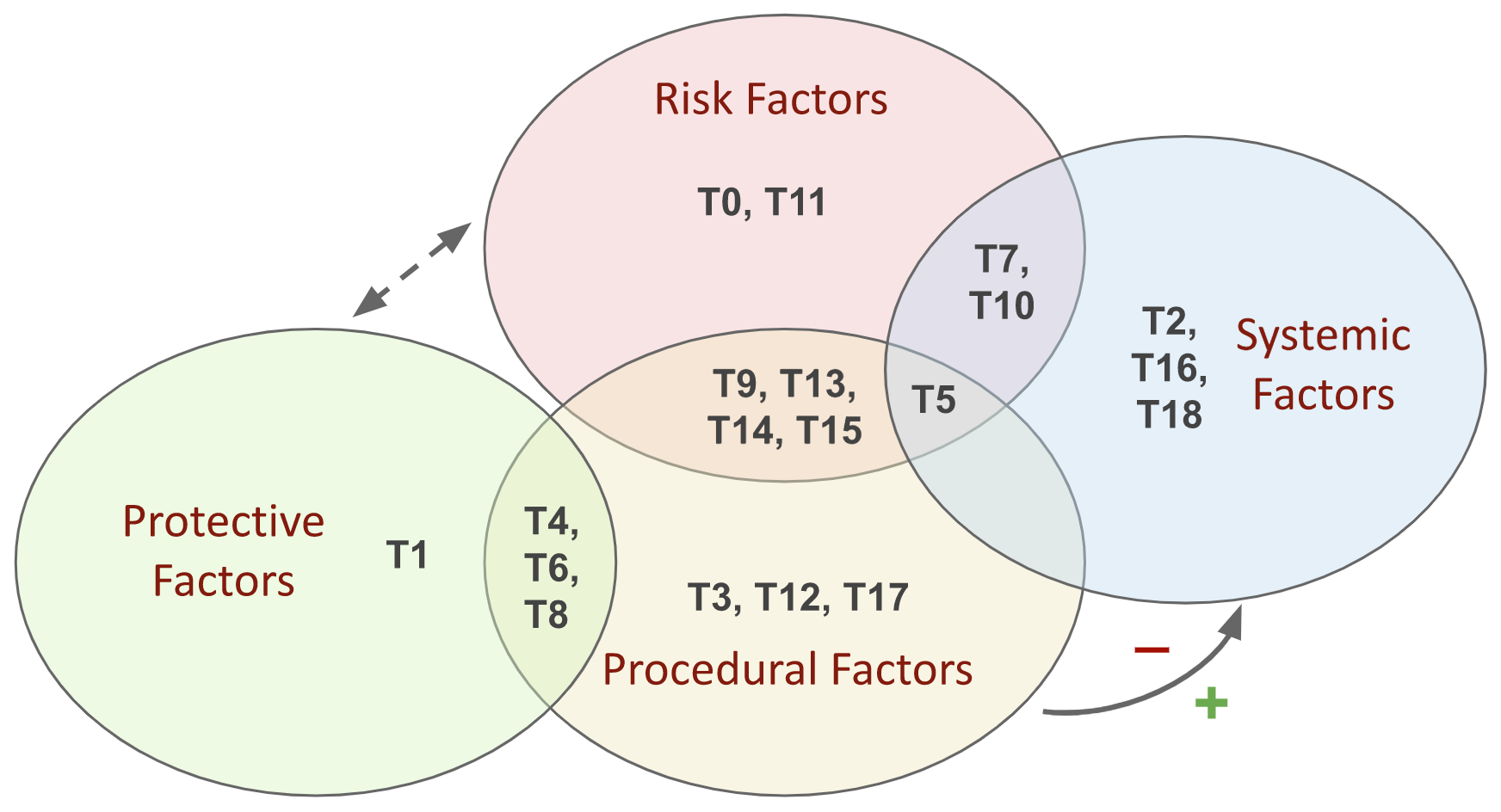}
    \caption{Axial Coding Paradigm: Relationship between Risk, Protective, Systemic, and Procedural Factors. Topics are labeled T0-T18 (see Table 3). The plus (+) and minus (-) signs between procedural and systemic factors mean that procedural factors can help mitigate systemic factors but can also amplify them. The dotted line between risk and protective factors means that there is a constant tension between building protective factors and alleviating risk factors.}
    \label{fig:axial_coding}
    \Description{Axial Coding Paradigm: Relationship between Risk, Protective, Systemic, and Procedural Factors. Topics are labeled T0-T18 (see Table 3). The plus (+) and minus (-) signs between procedural and systemic factors mean that procedural factors can help mitigate systemic factors but can also amplify them. The dotted line between risk and protective factors means that there is a constant tension between building protective factors and alleviating risk factors.}
\end{figure*}

While interpreting our topics based on top keywords and exemplar casenotes, we learned that there was an overlap between several key factors (i.e., risk, protective, systemic, procedural) where a topic could belong to more than one category. For instance, lack of adequate housing is a systemic factor, however, it poses a direct risk to families. Therefore, we conducted qualitative axial coding \cite{cresswell2012educational} to understand how these different factors were related to one another. We placed risk factors at the center of this process (i.e., the core phenomena) and then assessed how other factors influenced the core phenomena, procedures formulated to influence the core phenomena, or general strategies carried out as a response. Figure \ref{fig:axial_coding} depicts these inter-relationships between the four categories.

\subsubsection{Group Analysis of Topic Popularity Over Time (RQ3)}
Prior social work studies have found that the duration of time that families spend in the child-welfare system is related to the complexity of their respective cases \cite{pinna2015evidence, carnochan2013achieving}. Here, case complexity may depend on maltreatment type, financial need, substance abuse or health concerns, and the age or number of children. In light of heavy workloads carried by caseworkers and high turnover, agencies often group cases into high, medium, and low needs so caseworkers have more equal caseloads \cite{kothari2021retention}. Saxena et al. \cite{saxena2022unpacking} also found that time spent with CWS (i.e., number of interactions with CW services) can indicate case complexity. As such, we grouped families into three groups - Group 1 (low needs), Group 2 (medium needs), and Group 3 (high needs). Due to space considerations and to improve readability, we only focus on Group 3 in this study. Group 3 includes families with 40+ interactions with the agency. Next, we plotted topics from the trained CorEx topic model in Section \ref{section:CorEx} over time to understand which topics (i.e., risk, protective, systemic, and procedural factors) emerged as significant at different temporal points in a case. To accomplish this, we followed the methodology from Saxena et al. \cite{saxena2022unpacking}, where we concatenated casenotes for each family in each group and then chronologically arranged them. As these casenotes tracked the trajectory of CWS events, we then equally divided the casenotes into ten segments, so each segment had the same number of words \cite{saxena2022unpacking}). Equal segmentation of casenotes thus allowed us to create normalized segments of text that can track the \textbf{"Life of a Case"} for different families involved with CW services at the agency for differing lengths of time.

\begin{table}
\small
\centering      
\begin{tabular}{>{\raggedright}p{2.5cm}|>{\raggedright\arraybackslash}p{5cm}}  
\hline         
\textbf{Topics} & \textbf{Anchoring Words} \\ [0.5ex] 
\hline
Risk Factors & `neglect’, `violent', `anger', `drug', `criminal', `behavior'  \\   
Protective factors  & `encourage', `receptive', `protective', `family', `support', `care'  \\ 
Systemic factors & `rent', `job', `transport', `insurance', `medication', `resource'  \\ 
Procedural factors & `attorney', `court', `consent', `appointment', `evaluation', `voicemail' \\ [1ex]
\hline      
\end{tabular}
\caption{Anchor Words associated with Risk, Protective, Systemic, and Procedural Factors} 
\label{tab:anchorwords}  
\vspace{-0.5cm}
\end{table}


\section{Results}
In this section, we discuss our results organized by our three research questions. For the sake of readability, we present our semi-supervised topic model solution organized by our set of anchor words, i.e. - risk, protective, systemic, and procedural factors. Topics are grouped in Table 1 based on anchor words and labeled T0-T18.

\begin{table*}[]
\small
\begin{tabular}{|>{\raggedright}p{0.2cm}|>{\raggedright}p{2.4cm}|>{\raggedright}p{7.5cm}|>{\raggedright\arraybackslash}p{5.8cm}|}
\hline
\textbf{\#} & \textbf{Theme}  & \textbf{Topic} & \textbf{Unique keywords} \\
\hline
\multirow{4}{0.2cm}{1.} & \multirow{4}{2.4cm}{\textbf{Risk Factors that Impact Family Well-Being}} & \textbf{T0}: Substance Misuse and Mental Health Issues & \textcolor{mygray}{\textit{behavior, drug, anger, neglect, violent}} \\
                    &  & \textbf{T11}: Risks arising from Inability to Manage Child Behaviors & \textcolor{mygray}{\textit{mother, feels, expressed, child, frustrated, understand
}} \\
                    &  & \textbf{T7}: Risks arising from Environmental Factors or Past Trauma  & \textcolor{mygray}{\textit{choke, vented, trafficking, boyfriend, fight, steal}}  \\
                    &  & \textbf{T18}: Risks arising from High Medical Needs of Children    & \textcolor{mygray}{\textit{observes, documentation, provides, informs, appointment, recover}} \\
\hline
\multirow{4}{0.2cm}{2.} & \multirow{4}{2.4cm}{\textbf{Protective Factors that Impact Family Well-Being}}    & \textbf{T1}: Building Protective Factors in a Child's Ecosystem & \textcolor{mygray}{\textit{family, care, support, encourage, preservation, receptive, growing}} \\
                    &   & \textbf{T4}: Recording Parents' Progress during Supervised Visits & \textcolor{mygray}{\textit{kisses, burped, engaged, activity, redirected, attention}} \\
                    &   & \textbf{T6}: Addressing Parenting Challenges through Parenting Classes & \textcolor{mygray}{\textit{communicated, related, clarified, reiterated, enrichment, negative}} \\
                    &   & \textbf{T8}: Employing Parenting Techniques through Parenting Curriculum   & \textcolor{mygray}{\textit{parenting, curriculum, session, completed, chapter}} \\
\hline
\multirow{3}{0.2cm}{3.} & \multirow{3}{2.4cm}{\textbf{Systemic Factors that Impact Families and Decision-Making}} & \textbf{T2}: Critical Economic Resources Needed for a Stable Household  & \textcolor{mygray}{\textit{rent, resource, insurance, pay, medications, landlord}} \\
                    &   & \textbf{T10}: Unforeseeable environmental or systemic factors that augment risk &  \textcolor{mygray}{\textit{assisted, residence, supervised, settled, transition, issues}} \\
                    &   & \textbf{T16}: Access to Household Necessities through Public Assistance and Community Providers &  \textcolor{mygray}{\textit{communicated, collateral, home, pantry, bus}} \\
\hline
\multirow{8}{0.2cm}{4.} & \multirow{8}{2.4cm}{\textbf{Procedural Factors that Impact Decision-Making}} & \textbf{T3}: Legal Processes Associated with Child-Welfare Cases  & \textcolor{mygray}{\textit{appointment, court, consent, attorney, evaluation}}\\
                    &   & \textbf{T5}: Caseworkers' efforts towards Finding Services for Clients  &  \textcolor{mygray}{\textit{services, information, resources, health, mental, aoda}}\\
                    &   & \textbf{T9}: Risks Arising from Street-level Decisions and Time Constraints  &  \textcolor{mygray}{\textit{services, help, mother, housing, feels, provided, therapy, health}}\\
                    &   & \textbf{T12}: Managing Logistics Associated with Supervised Visitations, Classes, and Appointments  &  \textcolor{mygray}{\textit{visits, case, shared, information, discuss, check, aware}}\\
                    &   & \textbf{T13}: Relationship between Caseworkers and Families  &  \textcolor{mygray}{\textit{waited, voicemail, shared, frustrated, complaint, responded}}\\
                    &   & \textbf{T14}: Barriers Associated with Following Permanency Plan  &  \textcolor{mygray}{\textit{received, visits, missed, explained, called, reports, schedule}}\\
                    &   & \textbf{T15}: Conducting Home Visits and Safety Assessments  &  \textcolor{mygray}{\textit{observed, dressed, clean, free, marks, visible, injury}}\\
                    &   & \textbf{T17}: Recording and Sharing Details about Services, Classes, and Appointments  &  \textcolor{mygray}{\textit{room, arrived, played, visit, time, center}}\\
\hline
\end{tabular}
\caption{19 topic semi-supervised model solution organized by four sets of anchor words. Topics are labeled T0-T18.}
\vspace{-0.5cm}
\label{tab:six_themes}
\end{table*}

\subsection{Critical Factors Arising in Child-Welfare Cases that Impact Decision-Making}
In this section, we first discuss our results grouped by our sets of anchor words and explain the different risk, protective, systemic, and procedural factors that impact family well-being and the decision-making process. In exemplar casenotes below, FPS refers to the Family Preservation Specialist and CM refers to the Case Manager.

\subsubsection{Risk Factors that Impact Family Well-Being}
We first grouped our topics based on risk factors that arise in families and are noted by caseworkers in casenotes. Substance and/or alcohol misuse and mental health issues emerged as the most dominant risk factors (and also the most dominant topic) in the topic model solution. This finding aligns with prior literature that found that one-third to two-thirds of child abuse/neglect cases involve substance use disorder \cite{cwgateway_riskfactors, austin2020risk}. Reading through casenotes for this topic, and as depicted by the exemplar sentence below, we also learned that substance use disorder (SUD) generally overlaps with some mental health issues. That is, in cases where SUD was a concern, mental health services were frequently discussed alongside AODA (Alcohol and Other Drug Abuse) services. This finding is also consistent with prior literature \cite{cwgateway_sud, austin2020risk}.

\begin{myquote}
    \small{Topic 0 example: \textcolor{mygray}{FPS and Mr. BN discussed why Mr. BN has not been in contact with FPS. Mr. BN discussed having alcohol poisoning and explained that he felt embarrassed and did not want to talk with FPS. Mr. BN continued saying that he did not want people thinking that he was not interested in getting his daughter back. Mr. BN informed FPS that he is in anger management and AODA classes and shared location of the AODA and anger management classes.}}
\end{myquote}

\textit{Risks arising from environmental factors or past trauma} is the next dominant risk factor where domestic violence or intimate partner violence was consistently discussed in casenotes. Prior work has found that intimate partner violence, especially in the case of single parents can pose an ongoing risk to the family (i.e., the parent and child) \cite{taylorCA2009,austin2020risk}. This is also challenging for child-welfare workers because they are unable to include a significant other in case planning because they are not a biological parent and are not legally bound to the case \cite{core2018knowledge}. The exemplar sentence below depicts how intimate partner violence can create risky situations for the family. \textit{Inability to manage child behaviors} emerged as the next significant risk factor. These cases generally involve minor cases of neglect that can be addressed with parents developing proper intervention and disciplining skills that reinforce positive behaviors in children. Caseworkers, especially Family Preservation Specialists (FPS), work with parents through parenting classes offered at the agency. Another risk factor arises as a result of a lack of trust and a poor relationship between parents and caseworkers. We witnessed several examples of this in casenotes where parents believed that the caseworker was unable to handle the case or needed assistance. In the top exemplar casenote for this topic, parents say that they will be filing a complaint against the worker. Prior work has found that a healthy working relationship between parents and caseworkers is essential for achieving positive outcomes for families and ensuring that cases are not re-referred in the future \cite{carnochan2013achieving}. 

\begin{myquote}
    \small{Topic 7 example: \textcolor{mygray}{[Significant other] kicked down the apartment door. Downstairs neighbors were upset and apartment landlord stated that she had called the police and they gave her a number for reference. Affordable Rental was called to fix the door. FPS contacted for affordable rental to see if they could move Ms. AP [parent] to another apartment. They stated that they had no apartment available and she would have to wait until September to see if something becomes available for rent. FPS then transported Ms. AP to MPD [police department]. MPD was unable to locate the purse. EM [child] was asleep when everything had happened.
}}
\end{myquote}

\subsubsection{Building Protective Factors within Families}
The majority of cases of child maltreatment are cases of neglect that are referred to CWS due to deeper systemic issues such as lack of access to child care, lack of access to healthcare, and lack of affordable housing \cite{austin2020risk, carnochan2013achieving}. Such issues can be addressed by ensuring that parent(s) have additional caregiver support. As depicted by Topic 1 and the exemplar casenote below, caseworkers work with parents to get other family members (e.g., relatives, grandparents) involved so that the parents have additional support, especially during stressful circumstances in their lives. However, as depicted by the exemplar casenote below, working with extended family also requires caseworkers to address any familial conflicts that arise to ensure all parties align with the permanency plan for reunification and that parents have ongoing support. 

\begin{myquote}
    \small{Topic 1 example \label{hello}: \textcolor{mygray}{FPS [Family Preservation Specialist] and FSS [Family Support Specialist] walked into the home and FPS introduced FSS to MGM [Maternal Grandmother], MGF[Maternal Grandfather], EM [child], CH [child], and MA [birth mother]. While waiting for BK [birth father] to arrive, FPS And FSS sat at the kitchen table while EM colored pictures and CH played with toys and walked to and from the table. MGF and MGM were in and out of the room. Conversations about the negative behaviors the kids are experiencing happened, including MGM and MGF talking about the kids' tantrums, and being violent towards each other and the adults in the home. FPS acknowledged that these behaviors are hard to deal with and can be caused for various reasons. MGM and MGF's displeasure at BK's continuing to have visits was also discussed and FPS stated that at this time the court order must be followed and FPS cannot cancel supervised visitations.}}
\end{myquote}

Topics 6 and 8 depict parents' progress in parenting classes where they are addressing their parenting challenges and employing parenting techniques learned from the parenting curriculum. As depicted in the exemplar casenote below, the family preservation team works with parents through intervention tactics and parenting techniques on how to manage children's behavior and employ positive enforcement techniques to build up children's self-confidence and promote healthy habits and behaviors. Parents' progress in these classes is observed and documented in the casenotes and summaries are submitted to the court as part of documentation upon completion of classes/services. We witnessed several instances in casenotes where child-welfare involvement began due to the child(ren) engaging in risky behaviors (e.g., running away) and the parents' inability to manage such behaviors. However, it was interesting to note that even for cases where the identified target problem was children's behavior or actions, parents were still referred to several services (e.g., AODA services, therapy) and not just parenting classes.

\begin{myquote}
    \small{Topic 8 example: \textcolor{mygray}{FPS went over examples that MS [birth mother] could use. FPS suggested that MS use the activity `Panda and the Frog' with KJ [child] at the next visitation. FPS then went over tantrums and MS stated that KJ is at that age. FPS went over the stages of tantrums and techniques to use. FPS then talked about courage and building your child up. FPS provided MS with techniques on building her child up and examples of building your child down. FPS asked MS to give encouragement to her children once a day}}
\end{myquote}

\subsubsection{Systemic Factors that Impact Families}
Systemic factors have been extensively discussed in prior social work literature \cite{cwgateway_systemic, austin2020risk}. Environmental factors or community-level risk factors are other terms that are prominently used to describe characteristics that impact most families referred to CWS. These include access to affordable housing, employment, health services, public transportation, and public assistance, among others \cite{cwgateway_systemic}. As depicted by Topic 2 and the accompanying casenote below, such systemic issues can periodically arise within families, impact parents' ability to provide a stable household, and need to be addressed promptly and intuitively by both the parents and caseworkers to maintain stability. Here, the CW staff keeps information on community resource providers, service providers, and community centers that are able to help parents during unforeseeable circumstances such as loss of employment and housing and provide financial assistance (e.g., food stamps, travel vouchers, household necessities) that would offer some temporary relief to parents. An exemplar casenote for topic 10 depicted below shows how the caseworker and parent work together towards addressing their current housing problem.

\begin{myquote}
    \small{Topic 10 example: \textcolor{mygray}{FPS asked Mr. JP how his housing search was progressing. Mr. JP stated that the property management list that FPS provided was not as helpful as he hoped due to companies being out of state. FPS informed Mr. JP that FPS will provide him with more information when they meet next week. FPS also discussed with Mr. JP that [Community Provider] provides emergency financial assistance for housing once he secures a place. FPS asked Mr. JP if he had contacted City of [city name] Cribs for Kids program. Mr. JP stated that he had not contacted them but plans to. Mr. JP stated that he received a check in December and went on to state that court is requiring that he conduct x hours of volunteer work and x hours of application completion. FPS informed Mr. JP that FPS will contact him on Monday to schedule a meeting and provide additional housing and other resource information.}}
\end{myquote}

It is important to note the temporality of these systemic risk factors as they may arise and require the parents to seek temporary assistance through public programs, however, they are also collectively resolved by the parents and caseworkers and do not pose an ongoing risk to the family. The exemplar casenote below depicts how caseworkers and parents work together in resolving such risks within the restrictive framework of CWS.

\begin{myquote}
    \small{Topic 2 example: \textcolor{mygray}{MS [parent] stated she lost her job and will be without a job by the end of the month. MS stated this job was through a agency and she will try to find another job by the end of the month. MS stated she has already contacted the agency and they are helping her look for another job. FPS suggested that MS continue to work and continue to find a job until then. FPS also offered help with finding a job. MS asked FPS if she could help her with her student loans that are currently in default in order to not have her taxes garnished. FPS told MS they could focus on applying for the income-based repayment plan and see if she is able to be on a low cost plan.}}
\end{myquote}

\subsubsection{Procedural Factors that Impact Decision-Making}
Procedural factors refer to the legislative framework (or legal `procedures') that underscores the entire child-welfare process and must be followed by all involved parties. These processes also establish the constraints within which all decisions must be made. These include court proceedings, legal agreements, medical appointments and services, assessments and evaluations, and the signing of consent forms, among others. Topic 3 (exemplar casenote below), describes procedural factors associated with child-welfare cases. Topics 5 and 17 describe caseworkers' efforts in finding services for their clients and recording the details (i.e., time, location, frequency) of these services so they can be shared with other parties.

\begin{myquote}
    \small{Topic 3 example: \textcolor{mygray}{FPS met with Ms. BR [parent] one-on-one at the agency in meeting room Innovation. During the meeting, Ms. BR discussed her CPS case and criminal court proceedings. Ms. BR was able to complete the following consent forms: Family Preservation consent forms, RISE youth consents, and medical consents. Ms. BR informed FPS that the meeting had to be short as she currently has to meet IA [Initial Assessment] worker and supervisor IASW [Initial Assessment Social Worker] at [City] CPS regarding her new CPS case. FPS arranged for a meeting next week [date] at [x]pm as Ms. BR informed FPS this was the only time she was available to meet.}}
\end{myquote}

Topic 15 describes caseworkers' observations during home visits, supervised visits, and completion of quantitative assessments. We witnessed several assessments in the form of home safety assessments, mental health assessments, and parenting assessments being continually used by caseworkers throughout the life of cases. Caseworkers must follow DCF policy and periodically complete these assessments because it allows the department to collect consistent information about all cases.  

\begin{myquote}
    \small{Topic 15 example: \textcolor{mygray}{BK [parent] is refusing to work with any child welfare agency and stated he will not engage in any kind of services. In fact, BK was very upset and stated that once paternity is established, he will be filing for custody and not work with CPS. FPS kindly wanted to do an assessment for safety with BK but he refused.}}
\end{myquote}

Reading through the casenotes, we learned that caseworkers made continued attempts (via phone calls, emails, and in-person visits) to get in touch with all involved parties (i.e., parents, foster parents, relatives, etc.) to plan and schedule these visits (i.e., topic 12). Even though it is quite a mundane task, it requires significant ongoing effort. Topic 14 describes efforts made towards following the permanency plan as established under court conditions. Caseworkers are intimately involved in the parents' lives where they continually gather details from services, medical appointments, classes, and home visits as a means to provide information on case progress. However, this in-depth involvement of government officials in the lives of vulnerable families has been described as over-surveillance and policing of families involved in CWS where parents are \textit{recipients of support} but also \textit{subjects of regulation} \cite{roberts2007child, copeland2021s, abdurahman2021calculating}. In addition, tensions can arise between caseworkers and parents because of caseworkers' paradoxical role, i.e. - policing vs. supporting families \cite{roberts2007child}. This is also coupled with caseworkers carrying high caseloads as well as high turnover in the caseworker position such that cases are continually transferred between caseworkers \cite{shim2010factors, barak2006they}. As highlighted by the Topic 13 exemplar casenote below, such tensions can periodically arise when parents might feel that the caseworker is not doing enough to support them or the caseworker might believe that parents are not making enough progress towards the permanency plan. This casenote also provides a glimpse into how families are continually surveilled in their homes - the caseworker considers it necessary to record all interactions during a supervised visit and tells the family that they were not allowed to speak in their native language in his presence and that an interpreter would be needed.

\begin{myquote}
    \small{Topic 13 example: \textcolor{mygray}{Caregiver made comments about believing FPS was not able to handle the family and that they felt FPS needed assistance. The caregiver offered to speak to FPS supervisor on behalf of FPS to get more assistance. FPS declined this offer stating that FPS would speak to their supervisor. Caregiver stated that she was giving FPS a heads up that the family was going to file a complaint against FPS. The caregiver also questioned FPS about the grandmother giving children prescription medications. FPS told the family they were not allowed to speak in Spanish. FPS clarified that it is not that the family is not allowed but that an interpreter would be needed as FPS is not fluent in Spanish.}}
\end{myquote}


\subsection{Interplay between Risk, Protective, Systemic, and Procedural Factors}

We anticipated topics to emerge distinctly based on our anchoring of words associated with risk, protective, systemic, and procedural factors; however, we witnessed that several topics overlapped with other topics. For instance, we learned from reading the top exemplar casenotes that systemic factors can amplify risks within a family. In addition, procedural factors could help mitigate systemic factors and build protective factors, but also inadvertently amplify risk factors. Therefore, we conducted axial coding to understand how different factors were associated with one another and provide a visual representation in Figure 1. Below, we discuss the interplay between these factors.

\subsubsection{Procedural factors can help mitigate systemic factors but can also amplify them}
\label{Section5.2.1}
Caseworkers work closely with parents to address any systemic barriers (e.g., finding new employment, housing, etc.) that may inhibit case progress. They must intuitively come up with any solutions or even `half-fixes' that may temporarily resolve a stressful circumstance for a parent. Here, the agency may be employing an evidence-based practice model, however, several arbitrary decisions are still made on the ground by caseworkers. In the exemplar casenote below, the child displays signs of underlying trauma and needs professional help. However, the caseworker draws an arbitrary conclusion and tells the parent to ensure that their child is not watching violent videos or playing violent games even though there is no evidence to suggest that this leads to aggressive behaviors \cite{kuhn2019does}. It also puts the onus on the parent to `do something' in order to address the child's immediate behavior. It is much later in this case that the caseworker acknowledges the need for a psychological evaluation and the child seeing a school psychologist. Throughout our reading of casenotes, we witnessed several such instances where caseworkers engaged in defensive decision-making \cite{munro2019decision} where they formulated actions for parents to undertake just to be able to document that they were taking necessary steps and making decisions that purportedly addressed risks within the family.

\begin{myquote}
    \small{Example casenote: \textcolor{mygray}{ MS [parent] stated that the school called her again in regards to CJ [child] and his behavior at school. MS stated CJ was kicking and punched his teacher and assistant at school today. He has been more violent with other kids too as he purposely hit them in the head. CJ stated to his teachers he knows the head is the part that makes a person stay alive and that is the reason why he aims for people's head when he hits them. FPS asked MS to make sure CJ is not watching violent videos or playing violent games at night.}}
\end{myquote}

In addition, caseworkers may also engage in defensive decision-making when they anticipate risky situations or feel that they do not have enough expertise to effectively handle conflicts that might arise. In the exemplar casenote below, the caseworker feels uncomfortable managing jointly supervised visits for parents who have a history of domestic violence and bargains on an incomplete assessment to avoid these joint visits.

\begin{myquote}
    \small{Example casenote: \textcolor{mygray}{ TL [parent] stated that she spoke with her CM regarding HK [parent] attending joint visits. FPS explained that she would reach out to her CM regarding visits and  CM explained that HK would need to complete the assessment prior to his enrollment in the program. TL appeared upset that BK wouldn't be able to sit in today's visit and stated that he would have to sit in the car until her visit is over. FPS provided CM with an update on the conversation with TL. CM stated that she spoke with TL and explained that joint visits with HK wouldn't be appropriate given the father's mental health condition and history of DV. CM stated that she didn't feel comfortable having joint visits at this time.}}
\end{myquote}

Similarly, in another case (see casenote below), the parent tells the caseworker about a rodent infestation in their place and the caseworker helps the parent by speaking with the landlord. However, this is also followed by the caseworker conducting a home safety assessment that permanently records these new risks on the parent's case documentation.

\begin{myquote}
    \small{Example casenote: \textcolor{mygray}{MS [parent] discussed the issues that MS is having with her current landlord. MS stated that she thinks that there is a rodent infestation and that the landlord was not responding appropriately. MS stated that he was dragging his feet on an exterminator. FPS spoke with landlord to make sure he understands the urgency.}}
\end{myquote}

On the other hand, caseworkers within their capabilities, do help families address any arising risk or systemic factors (see example casenote below) associated with finding essential resources and getting access to public assistance. We also witnessed several instances where caseworkers helped parents create resumes for job applications, find new housing, apply for financial assistance, and get home essentials (e.g., beds for kids, clothing, toys, etc.). This underscores a need to understand the \textbf{\textit{why}} and \textbf{\textit{how}} street-level decisions are made by caseworkers within the restrictive legislative framework of CWS, as opposed to the broader focus on the service delivery model implemented at the agency.

\begin{myquote}
    \small{Example casenote: \textcolor{mygray}{LC [parent] shared that KC[child] was being bullied at school and asked if FPS could help her look into another middle school in the area for KC to attend. LC stated its to the point of her daughter having anxiety when she is getting ready for school. LC thought about enrolling her for online classes but wants that to be the last resort. FPS told LC that she will look into the list of different schools around the area that KC could possibly attend.}}
\end{myquote}

\subsubsection{Procedural Factors can Amplify Risk Factors}
Caseworkers are central to the child-welfare process and act as mediators between birth parents, the court system, and service providers \cite{ryan2006investigating, carnochan2013achieving}. That is, they bridge the administrative gap between legal processes established under court conditions (that the parents must conform to) and social work processes centered in helping families. These conflicting roles can create tensions between parents and caseworkers where caseworkers must help parents through services (e.g., therapy, parenting, domestic violence) but also police their actions to ensure that they are following court conditions for reunification \cite{copeland2021s, roberts2007child}. In the casenote below, the parents explain that they are more focused on finding stable housing and employment which is causing them to miss some supervised visits. Here, procedural factors add more stress to the lives of parents instead of helping them navigate child-welfare services. It is also important to note that the caseworker's primary concern here is receiving documentation about services so they are able to complete their procedural task. We witnessed similar tensions in other cases where the parent(s) shared that they were overwhelmed with several appointments for services and supervised visits throughout the week while trying to maintain full-time employment.

\begin{myquote}
    \small{Example casenote: \textcolor{mygray}{FPS discussed with CM as to whether the documentation is sufficient and how visits look moving forward. Mr/Ms KD said that FPS could call the service and they are not lying. FPS explained that she has name and number of the service and they have to have documentation [of services]. This affects their [parents] supervised visits with their son. Mr/Ms KD said they are pretty sure that the judge is to want that they have somewhere sufficient to live and this is a necessary step for the kids. They understand that they are missing visits but they also have to focus on the bigger goal [stable housing]. FPS asked if both are able to meet with her on Friday so we can get consents signed in order to verify employment and discuss visits moving forward.}}
\end{myquote}

In addition, caseworkers are mandated to follow court orders regarding who attends supervised visits. During court hearings, each parent's attorney advocates for their client's parental rights and ability to visit their children. However, as depicted in \hyperref[hello]{Topic 1 casenote}, inter-family conflicts can arise which pose an ongoing risk toward reunification efforts. Reading through the casenotes, we learned that there is a history of domestic violence in this case with one parent only peripherally involved in the children's life. Here, caseworkers must work with all involved parties even though uncertainty and conflicts persist within the family due to prior and ongoing risk factors. This further augments the overall uncertainty in decision-making because it is unclear whether familial conflicts would be resolved in the future so that both parents and relatives will provide caregiver support to each other.

\subsubsection{Interactions between Risk, Systemic, and Procedural Factors}
One critical aspect of CWS is to provide services (e.g., therapy, domestic violence, and Alcohol and Other Drug Abuse (AODA) classes) to parents to prevent future instances of child maltreatment. These services are agreed upon under court conditions, but it is up to caseworkers and parents to contact different service providers and find appointments. However, it can be challenging for both caseworkers and parents to find these services, especially accounting for parents' work schedules, supervised visitation appointments, and a lack of adequate service providers in the system. As depicted by the exemplar casenote for Topic 5 below, parents can find themselves waiting to hear back from service providers and may require assistance from caseworkers.

\begin{myquote}
    \small{Topic 5 casenote: \textcolor{mygray}{Family Peace [service provider] will work with YW [parent] on issues related to domestic violence. YW stated that she is still waiting to hear back from them as well as the agency to start therapy. FPS encouraged YW to call both agencies and let them know that she has been waiting to start services with them. FPS asked if she would give FPS permission to discuss her case with them.}}
\end{myquote}

Another critical issue here is the efficacy and consistency of services that are offered to parents and children \cite{mcmillen2006views, fedoravicius2008funneling, d2012parental}. Prior work has highlighted concerns associated with over-medication of children, overuse of psychotherapy, and inappropriate use of psychological testing \cite{mcmillen2006views}. As depicted in the exemplar casenote below, the parent expresses her concerns regarding another psychological evaluation for her child but is unable to exercise agency. Several states also employ a standardized service model or a "cookie cutter" approach where judges order therapy, services, and evaluations for all clients regardless of case circumstances \cite{mcmillen2006views}. Psychological evaluations, especially, act as catalysts and are often used as a "staple tool" by judges for the provision of mental health services \cite{fedoravicius2008funneling}. A program director at this agency confirmed that the Department of Children and Families (DCF) has used a cookie-cutter approach to services in the past where all parents had to complete the same set of services, i.e. - a standardized care approach was used instead of individualized care that recognizes target problems (e.g., mental health, drug use, domestic violence).


\begin{myquote}
    \small{Topic 5 casenote: \textcolor{mygray}{FPS asked KL [parent] if she wanted to have another psychological evaluation done due to him [child] drawing disturbing pictures of hurting other people. KL stated she is afraid that they will only put him on more medication.}}
\end{myquote}


\subsection{Temporal Dynamics between Factors through the Life of Cases}
Following the methodology from Saxena et al. (2022) \cite{saxena2022unpacking}, we grouped families into three groups based on their number of interactions with the agency. Below, we only focus on the group with the most interactions with CWS (i.e., Group 3) because these are the more complicated child-welfare cases where interactions between risk, protective, systemic, and procedural factors are more evident. Prior work has also highlighted that case complexity (e.g., type of maltreatment, age, number of children, need for financial assistance, drug abuse in the family) is directly associated with the time spent under the care of CWS \cite{pinna2015evidence, carnochan2013achieving}.

\subsubsection{Competing and Fluctuating Factors Lead to Uncertainty and Confounding Factors}

\begin{figure}[]
  \includegraphics[scale=0.38]{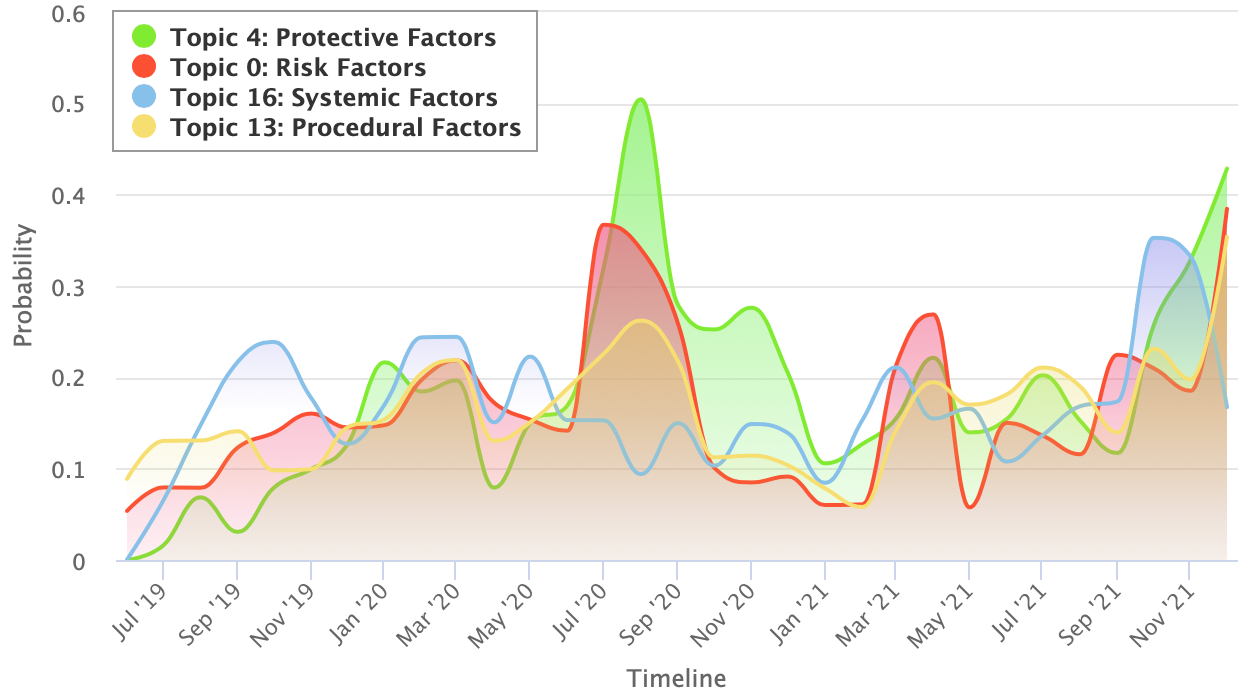}
  \caption{Relationship between Risk, Protective, Systemic, and Procedural Factors. Fluctuating and competing factors augment uncertainty and confound caseworkers' decision-making ability such that uncertainty about long-term family well-being often persists even at case closure}
  \label{fig:all_factors}
  \Description{Relationship between Risk, Protective, Systemic, and Procedural Factors. Fluctuating and competing factors augment uncertainty and confound caseworkers' decision-making ability such that uncertainty about long-term family well-being often persists even at case closure}
\end{figure}

As depicted in Figure \ref{fig:all_factors}, risk, protective, systemic, and procedural factors continually fluctuate and interplay with each other (i.e., changing topic probabilities over the life of cases). This may confound caseworkers' judgment and leads to uncertainty in decision-making because at any given time it is unclear what the trajectory of a case might look like. For instance, parents might be building protective factors where they now have additional caregiver support and learning parenting techniques that help them better manage child behaviors. However, systemic factors (e.g., loss of employment, housing, etc.) may also arise throughout the life of the case and pose risk to the permanency plan. As depicted in the previous section, procedural and systemic factors themselves may pose risks to families. Therefore, it is interesting to note that risk, systemic, and procedural factors oscillate together throughout the life of the case. This could be for two reasons - a discussion of needing services (i.e., \textit{risk}) is generally coupled with finding services (i.e., \textit{systemic}) and its association with the permanency plan (i.e., \textit{procedural}), and 2) caseworkers discuss any arising systemic or procedural factors followed by their impact on the family. Caseworkers also discuss the development of protective factors in great detail as depicted by the green trend. This is primarily the case because the majority of these casenotes come from the Family Preservation Team which works closely with parents through parenting services. In sum, a post-hoc analysis of these trends of competing factors shows that uncertainty about the final outcome of cases (i.e., reunification or placement in foster care) persists even at case closure where several cases are re-referred to CWS in the future \cite{klein2014neighborhood}. It also highlights that caseworkers continually face confounding factors (as a result of competing factors) in situ throughout the child-welfare process.


\subsubsection{Relationship between Different Protective Factors}

\begin{figure}[]
  \includegraphics[scale=0.38]{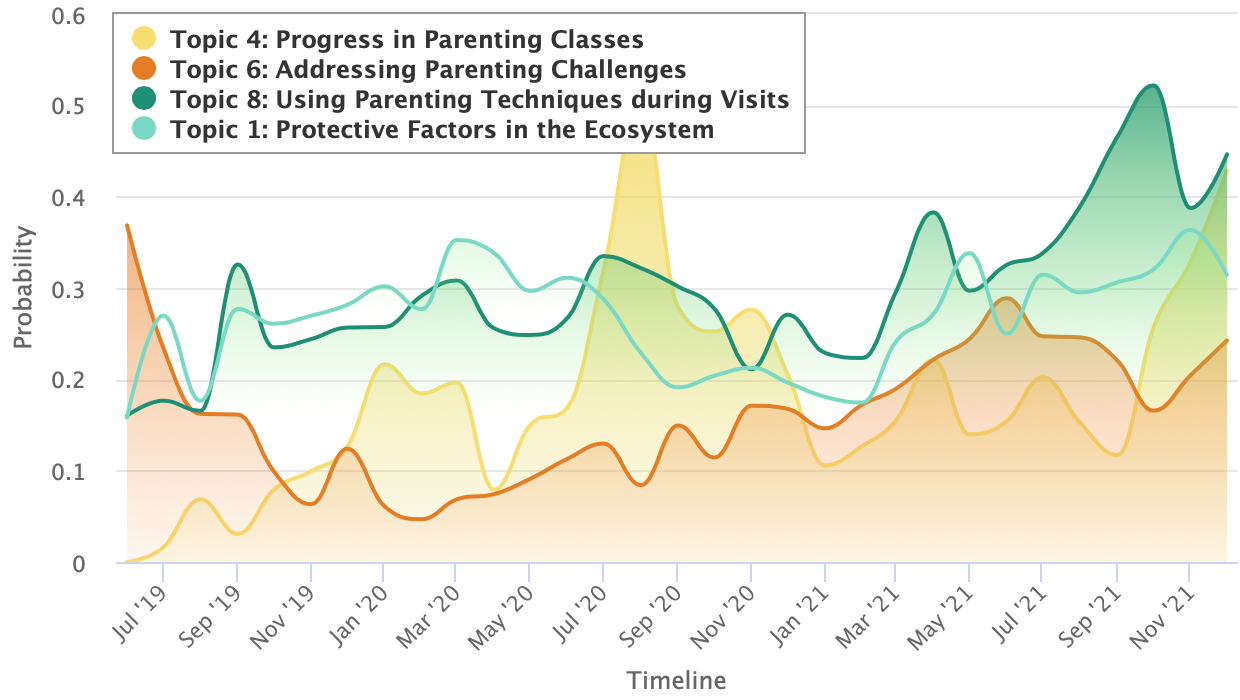}
  \vspace{-0.2cm}
  \caption{Relationship between different Protective Factors. Understanding the temporality and interplay between different protective factors can help assess long-term well-being outcomes for families.}
  \label{fig:protective}
  \Description{Relationship between different Protective Factors. Understanding the temporality and interplay between different protective factors can help assess long-term well-being outcomes for families.}
\end{figure}

Figure \ref{fig:protective} depicts trends in protective factors. Family Preservation Services works with parents in parenting classes and supervised visits to build protective factors. Teaching parents proper intervention and disciplining techniques helps address risks arising from the inability to manage child behaviors. Topic 1 highlights protective factors in a child's ecosystem by assessing their interactions within their social support system (i.e., parents, relatives, grandparents). We notice Topic 8 (i.e., employing parenting techniques during visits) follows a much similar trend as Topic 1. This may be the case because both topics inherently assess healthy and positive interactions between adults and children. On the other hand, Topic 6 describes caseworkers' conversations with parents on how they could be addressing parenting challenges as they work through the parenting curriculum. We expected this trend to be higher at the onset of cases but gradually diminish as parents develop protective skills and are recorded as observations in casenotes during parenting classes (i.e., Topic 4). This highlights the need to understand the temporality of such protective factors that help children and parents achieve positive developmental outcomes over time. This is often described as "resilience" in social work literature \cite{ungar2007contextual, lietz2011stories}. Resilience in children and parents is a result of interactions in their environment where caseworkers and other professionals can directly help shape this environment \cite{leadbeater2005resilience}. That is, an understanding of resilience can help assess which protective factors are pertinent for a family and would lead to better long-term outcomes.

\subsubsection{Relationship between Systemic Factors}

\begin{figure}
  \includegraphics[scale=0.38]{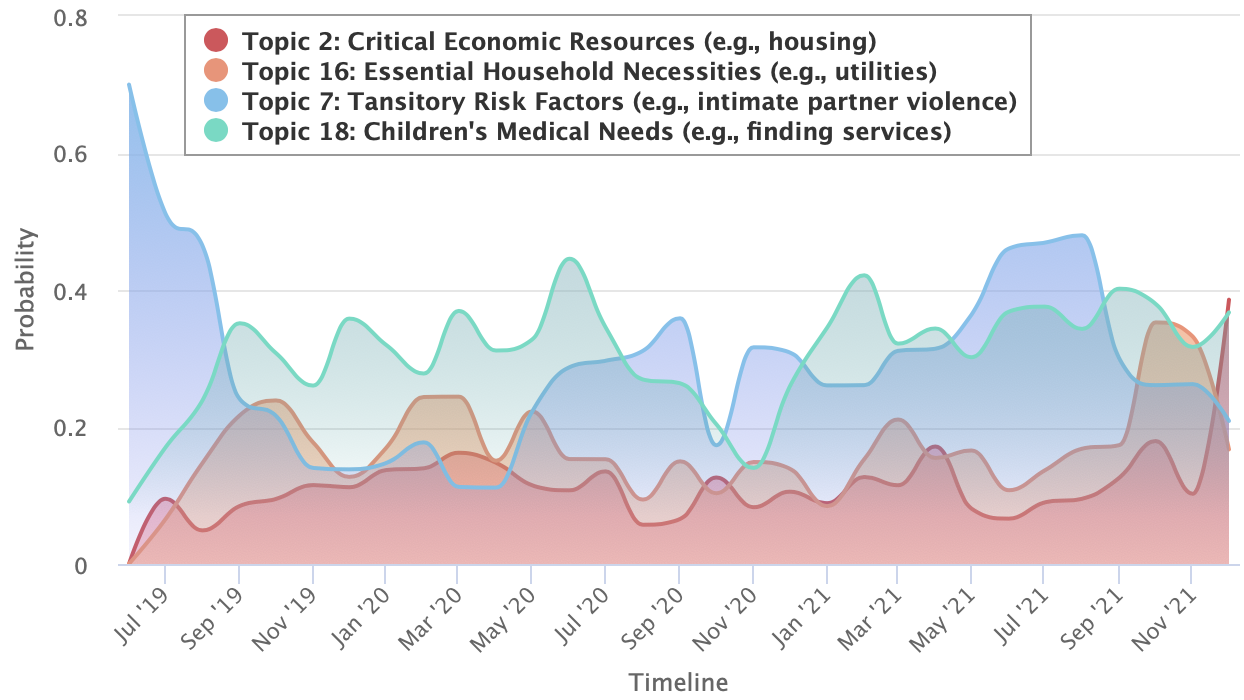}
  \vspace{-0.2cm}
  \caption{Relationship between different Systemic Factors. Both socioeconomic risk factors and transitory risk factors impact families. Caseworkers are able to address transitory risk with proper interventions but are unable to have a meaningful effect on socioeconomic risks.}
  \vspace{-0.3cm}
  \label{fig:systemic}
  \Description{Relationship between different Systemic Factors. Both socioeconomic risk factors and transitory risk factors impact families. Caseworkers are able to address transitory risk with proper interventions but are unable to have a meaningful effect on socioeconomic risks.}
\end{figure}

Figure \ref{fig:systemic} depicts trends in systemic factors. Topic 2 describes environmental and systemic factors that affect family well-being (e.g., employment, housing), and Topic 16 describes the essential household needs (e.g., food, clothing, utilities) as observed and recorded by caseworkers during home visits. In essence, both topics assess material resources necessary for maintaining a stable environment for children. This may explain why the trends for these two topics follow a much similar trajectory. On the other hand, Topic 7 describes emerging risk factors in a case due to unforeseeable circumstances such as intimate partner violence, medical needs arising from underlying trauma, and familial conflicts. We see a significant amount of fluctuation in this trend because unforeseeable systemic risks may arise but are also continually addressed through collaborative problem-solving between parents and caseworkers. Topic 18 discusses children's medical needs in terms of their medical appointments and medications. These needs as well as systemic barriers associated with meeting these needs (i.e., finding proper services, consistent mental health assessments, and medical appointments) are consistently recorded by caseworkers in their casenotes because this information needs to be shared among several involved parties. It is imperative to note here that structural economic issues (e.g., stable employment, safe and affordable housing, affordable health care) underscore the majority of child-welfare cases and involve poor and low-income families \cite{cwgateway_income}. These are the consistent risk factors (i.e., Topics 2 and 16) that impact most families. On the other hand, Topics 7 and 18 capture the transitory risk factors that the child-welfare staff is able to address with timely interventions. This underscores a need to understand both the socioeconomic risk factors that impact the majority of families as well as context-specific transitory risk factors specific to a family. Here, street-level interventions can help address some risks, however, systemic and policy-driven changes are equally necessary to improve social conditions that impact vulnerable and low-income communities.  

\begin{figure}[]
  \includegraphics[scale=0.38]{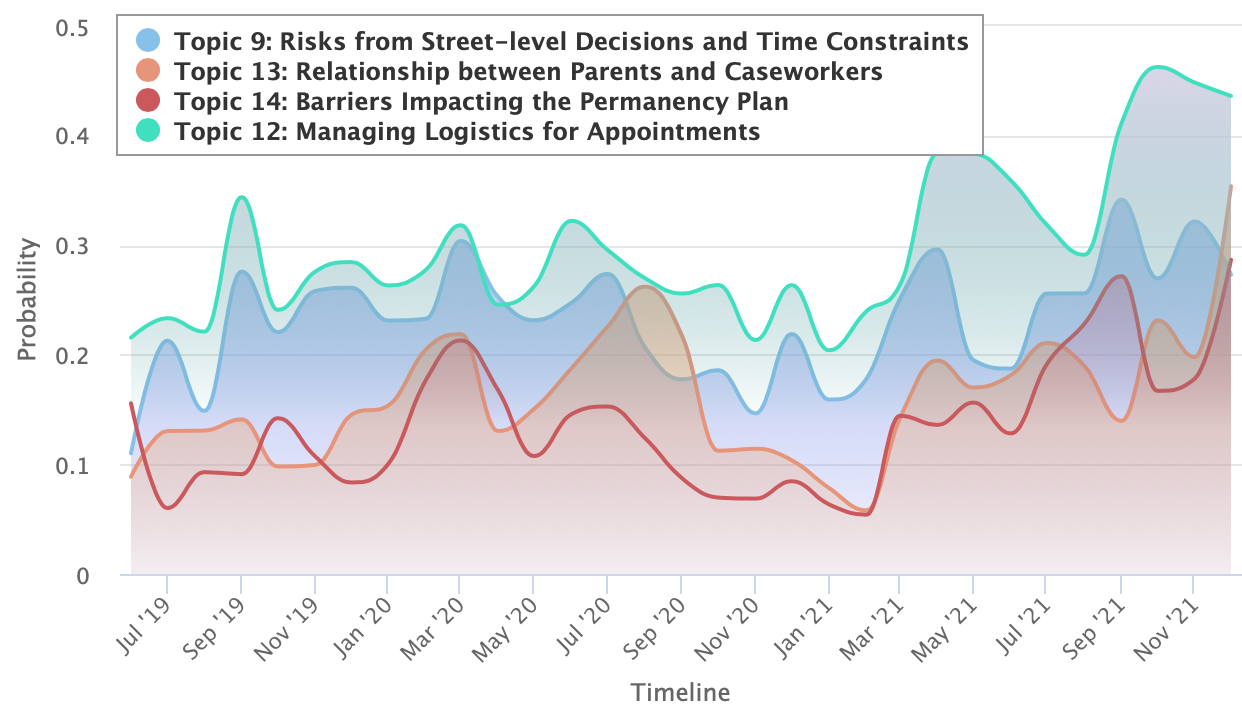}
  \vspace{-0.3cm}
  \caption{Relationship between Different Procedural Factors. Fluctuating procedural factors highlight tensions that arise between parents and caseworkers who must maintain a working relationship to make progress towards permanency.}
  \label{fig:procedural}
  \Description{Relationship between Different Procedural Factors. Fluctuating procedural factors highlight tensions that arise between parents and caseworkers who must maintain a working relationship to make progress towards permanency.}
  \vspace{-0.3cm}
\end{figure}

\subsubsection{Relationship between Different Procedural Factors}
Figure \ref{fig:procedural} depicts trends in procedural factors. Topic 13 describes strenuous relationships/interactions between caseworkers and parents. Child-welfare staff is legally mandated to follow a 15-month timeline which also establishes the permanency plan. That is, per the Adoption and Safe Families Act (ASFA), if parents do not fulfill all court conditions within 15 months, then parental rights must be terminated and child-welfare staff must find a more permanent placement for children in foster care. Here, caseworkers must work within this restrictive legislative framework to ensure that the permanency plan as established under court conditions is on track (i.e., Topic 14) where parents are completing court-ordered services, attending supervised visits, and working towards building a stable household, i.e., they must continually police the parents' actions to ensure progress towards permanency. Dorothy Roberts describes this as a dual and paradoxical role where caseworkers act as "investigators and helpers" and parents are both subjects of regulation and recipients of support \cite{roberts2007child}. These ongoing tensions between following the permanency plan and maintaining a working relationship with parents may explain why trends for these topics oscillate together. Moreover, Topic 12 describes scheduling and managing logistics around supervised visits and services. This trend is closely followed by Topic 9 which describes the risks emerging due to street-level decisions and time constraints. As noted in the previous section, parents in several cases shared that they were overwhelmed by the number of appointments and supervised visits while trying to maintain full-time employment and make necessary changes within their household. That is, procedural factors can themselves add risks to the stressful lives of parents who are fighting for reunification. Such risks arising due to the restrictive legislative framework of CWS cannot be quantified. It is also not possible to assess their long-term impact on families.

\section{Discussion}
Abebe et al. \cite{abebe2020roles} highlight that much of the computational research that focuses on fairness, bias, and accountability in machine learning systems continues to formulate “fair” technical solutions while treating problems that underscore the sociotechnical environment as fixed and fail to address deeper systemic and structural injustices. Through this study, we bring attention back to the \textit{sociotechnical} and highlight social problems in child-welfare and how these problems become embedded in algorithmic systems. Abebe et al. \cite{abebe2020roles} also formulate four roles or ways in which computational research can help address social problems. This study assumes the dual roles of \textit{\textbf{computing as rebuttal}} where we highlight the technical limitations and feasibility of predictive risk models (PRMs), and of \textit{\textbf{computing as synecdoche}} by uncovering systemic complexities and social problems in child-welfare that directly impact families.

\subsection{Rethinking "Risk" and the Underlying Data Collection Processes}
Our results bring into question how “risk” is formalized in the child-welfare system by drawing attention to the broader ecosystem of decision-making processes where systemic and procedural barriers can also create and amplify new risks posed to families. Prior research on algorithmic systems used in CWS has found that the majority of these systems define risk as a function of child and parent-related risk factors (e.g., parent’s involvement in drug and alcohol services, criminal justice, housing authority, etc.) \cite{saxena2020human, kawakami2022improving}, however, as our results show, the system itself can pose a significant amount of risk to families in regard to how protocols and practices (i.e., the legislative framework) are carried out on the street-level. This is further complicated by the fact that “risk of maltreatment” is poorly defined \cite{saxena2020human} (essentially comprising of three different outcomes – neglect, physical abuse, and sexual abuse), and this definition as well as criteria for investigating families can vary from one jurisdiction to another \cite{gambrill2000risk, dettlaff2011disentangling}. These investigations may result in substantiation of child maltreatment, and consequently, the case is brought into the system.

\textbf{Here, our results shed light on the data collection processes that ensue as parents are surveilled by caseworkers and mental health professionals.} We learned that caseworkers used several different screening tools and risk assessments that quantitatively capture risk factors during home visits, risk factors associated with children’s mental health, parents’ progress in parenting classes, as well as parents’ engagement and progress towards the permanency plan. The intent here is to collect as much information as possible and resolve any ambiguity resulting from missing information. However, this is problematic because CWS experiences a high turnover with a lack of well-trained caseworkers which leads to a lack of consistency in regard to how these assessments are completed \cite{shim2010factors, barak2006they}. Here, \textbf{caseworkers rely more on their impressions of the family in completing these assessments rather than expertise developed over time} \cite{copeland2021s, saxena2022train}.

In addition, algorithmic tools such as Allegheny Family Screening Tool (AFST) \cite{cheng2022child} and Eckerd Rapid Safety Feedback (ERSF) \cite{parker2022examining} use a family’s prior involvement in public and medical services to assess the risk of maltreatment through proxy outcomes of \textit{risk of re-referral} and \textit{placement in foster care}. However, as our results show, \textbf{parents lack agency in the process and must consent to assessments and information disclosures. They are unable to turn down services or classes that they might consider unnecessary.} Parents may also face repercussions and subsequently experience over-surveillance if they refuse psychological evaluations, drug tests, and/or additional services \cite{copeland2021s, roberts2007child}. As highlighted by Saxena et al. \cite{saxena2022train}, this refusal or disagreement with caseworkers might be captured under predictors such as “parents’ cooperation with the agency” – a significant predictor of risk per the WARM risk assessment. Services for parents and children are court-ordered where several states (including the state where this study was conducted) have employed a standardized service model in the past where parents in all cases were referred to a fixed set of services (e.g., parenting classes, psychological evaluations, Alcohol and Other Drug Abuse (AODA) services, etc.) regardless of case circumstances \cite{mcmillen2006views, fedoravicius2008funneling, d2012parental}. That is, \textbf{parents were enrolled in services that they did not necessarily need, and consequently, more data was collected about them through multi-institution partnerships between child-welfare agencies, service providers, and the court system.} Therefore, it is problematic for algorithms such as AFST and ERSF to use this cross-departmental data collected through power asymmetries because it further puts these families at a significantly higher risk of being re-investigated since their prior involvement with CWS renders them to receive “high-risk” predictions for future child maltreatment events. Our findings here, act as \textit{synecdoche} \cite{abebe2020roles} by making visible child-welfare practices and power asymmetries through which vulnerable low-income families are continually targeted by the system.

On the other hand, let us assume that designers and technologists developing algorithmic systems are able to adequately model for organizational context in terms of protocols, practices, resource constraints, and policies as well as make founded assumptions for a specific social context; then by extension, the developed system is no longer portable to a different jurisdiction or social context because child-welfare practice can vary significantly from one state to another. Selbst et al. \cite{selbst2019fairness} refer to this as the portability trap – “Failure to understand how repurposing algorithmic solutions designed for one social context may be misleading, inaccurate, or otherwise do harm when applied to a different context”. Here, we want to draw caution against child-welfare agencies acquiring algorithmic systems from private companies developed in one jurisdiction but sold and employed in several other jurisdictions \cite{eckerd, pap, mind-share, sas, augintel}. 

\subsection{Confounding Factors, Uncertainties, and Implications for Predictive Risk Models}
Prior literature in machine learning has discussed data and model uncertainties \cite{campagner2020three, gal2016dropout, kendall2018multi} and technical methods on how to mitigate these uncertainties that act as obstacles in the way of better predictive performance \cite{malinin2018predictive,ning2019optimization, kendall2017uncertainties}. On the other hand, HCI scholars have recommended that we engage with uncertainties as opportunities for human-centered design rather than treat them as obstacles \cite{soden2022modes, benjamin2021machine, paakkonen2020bureaucracy}. P{\"a}{\"a}kk{\"o}nen et al. \cite{paakkonen2020bureaucracy} note that “human discretionary power in algorithmic systems accumulates at locations where uncertainty about the operation of algorithms persists”. They further note that the design of algorithmic systems could benefit from identifying and cultivating important sources of uncertainties because it is at these sources that human discretion was most needed. 

\textbf{Our results in Section 5.3 move beyond data and model uncertainties and show how uncertainties can arise throughout the child-welfare pathway as a result of fluctuating factors} (i.e., risk, protective, systemic, and procedural factors) that continually interplay with each other and directly impact decision-making processes. Our results show that a parent may be developing protective factors through parenting services, however, transitory risk factors (e.g., loss of employment, housing, risks from intimate partner violence, etc.) may also periodically arise throughout the life of the case. In addition, systemic and procedural factors can themselves augment the risk posed to a family; however, at any given time there is a lack of clarity about their impact on the final outcome (i.e., reunification, adoption, or placement in foster care). \textbf{These competing and fluctuating factors confound caseworkers' decision-making and the situation is further aggravated by a lack of experienced caseworkers in the system \cite{shim2010factors, barak2006they}.}

This further brings into question our understanding of ecological risk and problematizes three core attributes regarding how risk is modeled in algorithms - 1) different risk factors present in a case are modeled as static variables, however, as our results show, transitory risk factors may arise but are also addressed collectively by caseworkers and parents. Here, \textbf{risk as a static construct is inherently biased because no temporal point estimate of risk taken at any given point in the child-welfare process offers a true picture of occurrences within the case}, 2) the baseline assumption underscoring predictive risk modeling is that risk within a family is likely to escalate if no interventions are made \cite{saxena2020human}. This leads to excessive CWS interventions and over-surveillance of vulnerable families \cite{copeland2021s, abdurahman2021calculating}. In addition, as our results show, the trends in risk factors oscillate throughout the life of cases where risk factors may arise but are also addressed. \textbf{Ignoring the temporality of different risk factors and treating them as static variables leads to elevated risk scores for families and excessive investigations}, and 3) prior work has established that empirical knowledge in child-welfare is quite limited and there is still a significant amount of debate regarding which risk factors (when taken together) lead to the accumulation of risk and which protective factors help mediate these risks \cite{saxena2020human, gambrill2001need, shlonsky2005next}. As our results in Section 5.3 suggest, different factors mediate the effects of each other throughout the child-welfare process. \textbf{Without understanding and embedding empirical knowledge of interaction effects within predictive risk models such as AFST \cite{cheng2022child} and ERSF \cite{parker2022examining}, risk predictions are likely to be elevated and biased.} As shown by a recent study conducted by Cheng and Stapleton et al. \cite{cheng2022child}, call screeners helped reduce racial disparities in AFST-predicted decisions by using their contextual knowledge of cases to override erroneous decisions. That is, an algorithm designed to bring neutrality and objectivity to the decision-making process is itself producing racially biased predictions. Our findings here, act as \textit{rebuttal} \cite{abebe2020roles} by highlighting the limitations of predictive risk models and the core assumptions about risk factors that make predicted outcomes unfeasible.

In addition, our results also draw attention to how seemingly mundane tasks carried out by caseworkers such as continued attempts to contact birth parents, foster parents, and relatives to schedule supervised visits and services can pose risks to the 15-month timeline of the permanency plan \cite{miller2009you} because it significantly impacts caseworkers' ability to work with parents and meet goals for completing set hours of visitations and services. This risk posed to families is hard to estimate but continually impacts street-level decision-making. It also highlights invisible patterns of labor that are only documented in casenotes and cannot be encapsulated by quantitative risk assessments.

\subsection{Implications for Computational Narrative Analysis and NLP-based Systems}
Selbst et al. \cite{selbst2019fairness} note that social context is often abstracted out so that machine learning tools can be applied to any given domain and evaluated based on predictive performance (i.e., \textit{the algorithmic frame}). Here, fair machine learning researchers may further expand upon this approach to investigate ML system’s inputs and outputs (i.e., \textit{the data frame}), however, this is still an attempt to formulate mathematical notions of “fairness” and “bias” and continues to abstract out the broader social context within which the system is situated and interacts within organizational pressures, systemic constraints, and with a variety of stakeholders. Consequently, authors formulate the \textit{sociotechnical frame} which recognizes that an ML model is simply a subsystem within a broader sociotechnical system where drawing stakeholders and institutions into the abstraction boundary allow us to investigate complex interactions. Through this study, we used computational narrative analysis to draw attention back to the sociotechnical frame and highlighted the complicated interactions between caseworkers and families, and brought into focus the critical structural issues within CWS. 

Computational methods such as unsupervised and semi-superv -ised topic modeling \cite{boyd2017applications,gallagher2017anchored} facilitate a qualitative exploration of casenotes and allow us to understand street-level practices, systemic constraints, power asymmetries, as well as temporal dynamics between different factors. Prior work has hypothesized that using text-based narratives within risk assessment algorithms may offer more holistic and fair predictions by filling in the gaps of quantitative risk predictors \cite{saxena2020human}. However, we want to draw caution against this approach as tech companies are beginning to pitch NLP-based systems to human services agencies and are being acquired by several agencies across the United States \cite{augintel}. Here, it is important to note whose values become embedded in these systems \cite{birhane2022values} and which (and whose) resources are directed towards these initiatives in an overburdened and underfunded system \cite{gilroy2018critical}. Through our reading of casenotes and data analysis, we recognized several limitations associated with caseworkers' narratives.

First, as previously noted in the Methods section, we manually analyzed several data sources to assess which casenotes contained detailed and credible information about interactions between parents and caseworkers. We settled upon casenotes written by the Family Preservation Team because they work closely with families throughout the process and understand the risks and needs associated with each family. On the other hand, \textbf{casenotes written by the initial assessment (IA) or investigative caseworkers carried information about \textit{perceived risks} and the caseworker's impression of the family} because not enough information is available (and at times, contradicting facts are present) at the onset of a case. Second, \textbf{even though the agency has established rigorous standards on documentation, there is variability in the writing of casenotes} where some caseworkers captured details about children's demeanor during transportation, supervised visits, and medical appointments while other caseworkers only wrote pertinent details (e.g., child's response to parenting techniques, medication schedule created at a medical appointment, transportation logistics). Third, \textbf{inexperienced caseworkers are known to engage in defensive decision-making where they might omit their mistakes from casenotes} \cite{munro2019decision}. As highlighted in the second exemplar casenote in {\hyperref[Section 5.2.1]{Section 5.2.1}}, the caseworker prioritized their own comfort over conducting joint supervised visits leading to the frustration of parents. This interaction was only uncovered because the Family Preservation Specialist wrote about it in their casenote entry. Here, collaboratively written documentation offers some accountability but we hypothesize that there may be several other such instances where caseworkers' actions went unchecked and undocumented. Fourth, qualitative exploration of casenotes driven by our computational approach allows us to understand the power asymmetries that both parents and caseworkers experience in the child-welfare process, however, \textbf{the contextual knowledge derived from casenotes can easily be stripped and instead exploited once quantified to be used in downstream tasks in NLP-based systems}. Fifth, as noted in the previous two sections, \textbf{a discussion of risk in casenotes does not necessarily mean that it is a persistent danger impacting family well-being}. It may simply be a noteworthy event that a caseworker recorded at that point in time (e.g., a child being bullied at school).

These five points are crucial for developing an understanding of this complex sociotechnical system and are especially important as it pertains to natural language processing. Recent studies in NLP have examined text datasets and seed lexicons and found that social hierarchies and racial, cultural, and cognitive biases can become embedded in and amplified by NLP systems and lead to significant disparities in downstream tasks \cite{antoniakbad, sun19, caliskan17, blodgett20, anjalie21}. Our findings here, act as a \textit{rebuttal} by highlighting the limitations of casenotes as a data source for downstream NLP tasks and act as \textit{synecdoche} by making visible the structural issues in child-welfare that become embedded in casenotes. Alternately, an upstream approach (i.e., - the corpus itself becomes an object of the study {\cite{antoniak2022modeling}}) as adopted by this study can help uncover contextual street-level interactions, critical factors that are hard to quantify, and uncertainties and confounding factors that offer a more comprehensive view of the decision-making ecosystem. In addition, an upstream approach allowed us to uncover empirical evidence about how marginalized communities face systemic injustices in child-welfare. 


\section{Limitations and Future Work}
This study conducts a computational narrative analysis of casenotes at one child-welfare agency in a midwestern state in the United States and uncovers several factors and street-level interactions that impact decision-making and family well-being. However, this study has some limitations that create opportunities for researchers to further expand upon this body of work. First, child-welfare practice can significantly vary from one state to another in terms of criteria for investigations and policies and protocols that all parties are mandated to follow. Similar analyses conducted in other jurisdictions would reveal hyperlocal and context-specific experiences of caseworkers and families in those regions. Second, this study only uncovers interactions and street-level decisions through the perspective of caseworkers and may omit or underplay the oppression, surveillance, and coercion experienced by many families and cannot reveal structural power dynamics that fundamentally underpin child-welfare interactions \cite{copeland2021s, roberts2007child}. These interactions are socially situated where parents are likely to experience the same events differently. Here, it is important to understand the perspective of affected communities (i.e., foster children, parents, and foster parents) about whom decisions are being made. For instance, a recent study conducted by Stapleton et al. \cite{stapleton2022imagining} found that parents considered CWS to be punitive and unsupportive and instead wanted systems that would help them fight against CPS as well as evaluate CPS and the caseworkers themselves. Future research should continue to focus on uncovering street-level complexities within this complicated sociotechnical environment through the perspective of families and caseworkers. Finally, this study takes an upstream, corpus-focused approach (i.e., the corpus itself is the object of the study) where we sought to understand dynamic and transitory factors embedded in caseworkers' narratives that impact decision-making and family well-being. However, this requires a re-analysis and experimentation of different NLP techniques to focus on topic-specific corpora such as the corpus used in this study. That is, we recommend that researchers conducting similar analyses in various public sector domains experiment with and compare different NLP methods to assess which methods help uncover latent signals in the corpus as well as highlight limitations in the corpus itself.


\section{Conclusion}
We conducted a computational narrative analysis using Correlation Explanation (CorEx) \cite{gallagher2017anchored}, a semi-supervised topic modeling approach that allows us to incorporate domain knowledge in the form of anchor words. Using the socioecological model of health and development \cite{austin2020risk} as our theoretical lens, we incorporated domain knowledge about risks, protective, systemic, and procedural factors that impact decision-making and family well-being. We provide empirical evidence that the child-welfare system itself poses a significant amount of risk to the families that it is expected to serve. We show how parents are over-surveilled in the system, the lack of agency they experience in the child-welfare process, and problematize the data collection processes that take place as a result of this power asymmetry. We complicate the use of predictive risk models that treat risk factors as static constructs by highlighting the multiplicity and temporality of different risk factors that arise throughout the child-welfare pathway. Finally, we draw caution against using casenotes in NLP-based systems by highlighting the limitations and biases embedded within this data source.

\begin{acks}
This research was supported by the NSERC Discovery Early Career Researcher Award (\#04570) and the University of Toronto’s Human-AI Interaction Summer Research School (https://www.thai-rs.com). Any opinions, findings, conclusions, or recommendations expressed in this material are those of the authors and do not necessarily reflect the views of our sponsors or community partners. We would like to thank our collaborators at the child-welfare agency, \textit{Wellpoint Care Network}, for sharing the casenotes dataset and providing valuable feedback on our data analysis process. Finally, we are also thankful for the anonymous reviewers whose suggestions and comments helped improve the quality of this manuscript.
\end{acks}


\bibliographystyle{ACM-Reference-Format}
\bibliography{main}

\appendix

\end{document}